\documentclass{article}

\usepackage{amssymb,amsfonts,amsmath,stmaryrd}
\usepackage{cite,enumerate,float,indentfirst}
\usepackage{color}
\usepackage{comment}
\usepackage{url}

\def\be{\begin{eqnarray}}
\def\ee{\end{eqnarray}}
\def\nn{\nonumber}
\def\MD{\hbox{Md}}
\def\dMD{{^\vee\hbox{Md}}}
\def\qDv{{^\vee\!\hbox{qD}}}
\def\qD{\hbox{qD}}

\usepackage{ dsfont } 

\def\be{\begin{eqnarray}}
\def\ee{\end{eqnarray}}
\def\nn{\nonumber}

\newcommand{\beq}{\begin{equation}}
\newcommand{\eeq}{\end{equation}}
\newcommand{\beqa}{\begin{eqnarray}}
\newcommand{\eeqa}{\end{eqnarray}}

\definecolor{red}{rgb}{1,0,0}
\definecolor{orange}{rgb}{1,0.5,0}
\definecolor{violet}{rgb}{0.7,0,1}

\definecolor{red}{rgb}{1,0,0}
\definecolor{orange}{rgb}{1,0.5,0}
\definecolor{violet}{rgb}{0.7,0,1}

\hoffset= - 1.0in         

\begin{document}

\title{\vspace{1.5cm}\bf
On Refined Vogel's universality
}

\author{
Liudmila Bishler$^{a,b,c,}$\footnote{bishlerlv@lebedev.ru} and
Andrei Mironov$^{a,b,c,}$\footnote{mironov@lpi.ru,mironov@itep.ru}
}

\date{ }

\maketitle

\vspace{-6cm}

\begin{center}
  \hfill FIAN/TD-07/25\\
  \hfill ITEP/TH-13/25\\
  \hfill IITP/TH-11/25
\end{center}

\vspace{4.5cm}

\begin{center}
	$^a$ {\small {\it Lebedev Physics Institute, Moscow 119991, Russia}}\\
	$^b$ {\small {\it ITEP, Moscow 117218, Russia}}\\
	$^c$ {\small {\it Institute for Information Transmission Problems, Moscow 127994, Russia}}
\end{center}

\vspace{.1cm}

\begin{abstract}
In accordance with P. Vogel, a set of algebra structures in Chern-Simons theory can be made universal, independent of a particular family of simple Lie algebras. In particular, this means that various quantities in the adjoint representations of these simple Lie algebras such as dimensions and quantum dimensions, Racah coefficients, etc. are simple rational functions of two parameters on Vogel's plane, giving three lines associated with $sl$, $so/sp$ and exceptional algebras correspondingly. By analyzing the partition function of refined of Chern-Simons theory, it was suggested earlier that the refinement may preserve the universality for simply laced algebras. Here we support this conjecture by analysing the Macdonald dimensions, i.e. values of Macdonald polynomials at $q^\rho$, where $\rho$ is the Weyl vector: there is a universality formula that describes these dimensions for the simply laced algebras as a function on the Vogel's plane.
\end{abstract}

\bigskip

\newcommand\smallpar[1]{
  \noindent $\bullet$ \textbf{#1}
}

\section{Introduction}

For more than 2 decades, P.\,Vogel \cite{Vogel95,Vogel99} worked on the construction of a Universal Lie algebra which would incorporate all simple Lie algebras. This idea has not found its ultimate realization as an actual algebra; however, a concept of universal quantity has been created. This concept implies that various quantities in Chern-Simons theory with gauge group $G$ associated with structures of the Lie algebra $g$ have a universal description, i.e. can be described as rational functions of two Vogel parameters. In fact, in the plane of these two parameters, one naturally distinguishes three straight lines: those associated with $sl$, $so/sp$ and exceptional algebras.

The universal quantities has to do with algebras, but not with their representations. This is quite natural, since the structures of representations of distinct simple Lie algebras are too much different. Thus, one expects the universal quantities have to be associated only with the adjoint representation and its descendants. Indeed, the typical universal quantities are: the Chern-Simons partition function \cite{MkrtVes12,Mkrt13,KreflMkrt,M2}, the dimension \cite{Vogel99} and quantum dimension \cite{Westbury03,MkrtQDims} of the adjoint representation, eigenvalues of the second and higher Casimir operators \cite{LandMan06,MkrtSergVes,ManeIsaevKrivMkrt, IsaevProv, IsaevKriv,IsaevKrivProv}, volume of simple Lie groups \cite{KreflMkrt},
the HOMFLY-PT knot/link polynomial colored with adjoint representation \cite{MMM,MM} and the Racah matrix involving the adjoint representation and its descendants \cite{MM,ManeIsaevKrivMkrt, IsaevProv, IsaevKriv,IsaevKrivProv} (see also \cite{KLS}).

In description, it is simpler to use three Vogel's parameters giving the projective plane: $\mathfrak{a}$, $\mathfrak{b}$ and $\mathfrak{c}$ such that one can scale all of them at once with an arbitrary constant. One usually chooses one of the parameters, $\mathfrak{a}$ to be -2. Note that these parameters are usually denoted as $\alpha$, $\beta$ and $\gamma$, however, we denote them differently, since $\alpha$ is used in order to denote roots of the root system.
The Vogel's parameters for simple Lie algebras are listed in Table \ref{vogelparm}.
\begin{table}[!ht]
\centering
\begin{tabular}{|c|c|c|c|c|c|}
\hline
Root system & Lie algebra & $\mathfrak{a}$ & $\mathfrak{b}$ & $\mathfrak{c}$ & $\mathfrak{t} = \mathfrak{a}+\mathfrak{b}+\mathfrak{c}$ \\
\hline
$A_n$ & ${sl}_{n+1}$ & $-2$ & $2$ & $n+1$ & $n+1$ \\
$B_n$ & ${so}_{2n+1}$ & $-2$ & $4$ & $2n-3$ & $2n-1$ \\
$C_n$ & ${sp}_{2n}$ & $-2$ & $1$ & $n+2$ & $n+1$ \\
$D_n$ & ${so}_{2n}$ & $-2$ & $4$ & $2n-4$ & $2n-2$ \\
$G_2$ & ${g}_2$ & $-2$ & $\frac{10}{3}$ & $\frac{8}{3}$ & $4$ \\
$F_4$ & ${f}_4$ & $-2$ & $5$ & $6$ & $9$ \\
$E_6$ & ${e}_6$ & $-2$ & $6$ & $8$ & $12$ \\
$E_7$ & ${e}_7$ & $-2$ & $8$ & $12$ & $18$ \\
$E_8$ & ${e}_8$ & $-2$ & $12$ & $20$ & $30$ \\
\hline
\end{tabular}
\caption{Vogel's parameters}
\label{vogelparm}
\end{table}

There were also attempts to extend the notion of universality to the refined Chern-Simons theory \cite{KS,AM1,Mane}. In practice, it was realized in \cite{KS} for the classical Lie algebras and in \cite{AM1,Mane} for the exceptional algebras that the partition function of the refined Chern-Simons theory is universal only for the simply laced algebras. Note that there is another universal formula, which unifies knot hyperpolynomials for the root systems $A_1$, $A_2$, $D_4$, $E_6$, $E_7$, $E_8$ \cite{ChE}, though it could potentially include also $G_2$, $F_4$ \cite{DG}, but does not include, because these two are not simply laced. In this brief note we extend the analysis to the case of Macdonald dimensions, and realize that they are also universal only in the simply laced case. The technical reason for this is that, in the refined case, only the dual Macdonald dimensions are factorized, and they coincide with the Macdonald dimensions only in the simply laced case.

\paragraph{Notation and comments.}
\begin{itemize}
\item To denote the roots of the root system $R$, we use the letter $\alpha$, and we denote the set of all positive roots as $R_+$.
\item
In variance with the original work by I. Macdonald \cite{Mac} and subsequent papers on the subject \cite{MacConj,CherednikConj,CherednikDAHA,Koorn},
we use symmetric quantum numbers, which allows us to present the results in a shorter and more elegant form: in our notation, the Macdonald polynomial depends on the squares of the original Macdonald parameters:
\begin{equation}
    q \,\rightarrow \, q^2, \,\, t \,\rightarrow \, t^2,\,\,  t_{\alpha}^2 \,\rightarrow \, t_{\alpha}^2
\end{equation}
and the symmetric bracket and quantum numbers are defined to be
\begin{equation}
    \{x\} = x-x^{-1}, \quad \{x\}_{_+}=x+x^{-1}, \quad[n]_q = \frac{q^n-q^{-n}}{q-q^{-1}}, \quad [n]_t = \frac{t^n-t^{-n}}{t-t^{-1}}.
\end{equation}
\item When we say that some symmetric polynomial that depends on variables $x_1, \dots, x_n$ is taken at the point $q^{\rho}$, where $\rho = (\rho_1, \dots,\rho_n)$ is the Weyl vector, we mean that one should make the substitution in this symmetric polynomial
\begin{equation}
    x_i = q^{\rho_i}.
\end{equation}
The number of variables $x_i$ and the length of the Weyl vector coincide and are equal to the dimension of the Euclidean space where the root lattice embedded.
\item We use the minimal normalization of the roots, which means that the length of the longest root $(\alpha_l,\alpha_l) = 2$.
\end{itemize}

\section{Universality in Chern-Simons theory}

We start with unrefined Chern-Simons theory.
The first universal formula is the formula of dimensions of adjoint representations \cite{Vogel95} of all simple Lie algebras:
\begin{equation}
\boxed{
  D_{\text{Adj}}(\mathfrak{a},\mathfrak{b},\mathfrak{c})  = {(\mathfrak{a} - 2\mathfrak{t})(\mathfrak{b} - 2\mathfrak{t})(\mathfrak{c} - 2\mathfrak{t})\over\mathfrak{a}\mathfrak{b}\mathfrak{c}}
  }
\end{equation}

\vspace{10pt}
The next natural quantity to consider is the quantum dimension $\qD_{\lambda}$. In Chern-Simons theory, {\bf quantum dimension} is the Wilson averages of unknotted loop (unknot invariant in knot theory). In representation theory, the quantum dimension is the character of the Lie algebras at a special point $x = q^{2\rho}$, where $\rho$ is the Weyl vector, which is a half sum of the positive roots of the corresponding root system $R$:
\begin{equation}
    \rho = \frac{1}{2}\sum_{\alpha\in R_+} \alpha.
\end{equation}
Characters of the Lie algebra given by the root system $R$, $\chi^{R}_{\lambda}$ do factorize \cite[expr. 13.170]{DiFr} at the Weyl vector:
\begin{equation}\label{qd}
    \qD_{\lambda}^{R}:= \chi^{R}_{\lambda} \left(x = q^{2\rho} \right) =\prod_{\alpha\in R_+} \frac{[(\alpha,\lambda+\rho)]_q}{[(\alpha,\rho)]_q}
\end{equation}
where $(\cdot,\cdot)$ is the scalar product in the Euclidean space where the root system is embedded.

The quantum dimensions were universalized in the adjoint representation \cite{Westbury03}, and they are equal to
\begin{align}
\boxed{
        \qD_{\text{Adj}}(\mathfrak{a},\mathfrak{b},\mathfrak{c}) ={[\mathfrak{a}/2 - \mathfrak{t}]_q[\mathfrak{b}/2 - \mathfrak{t}]_q[\mathfrak{c}/2 - \mathfrak{t}]_q\over[\mathfrak{a}/2]_q[\mathfrak{b}/2]_q[\mathfrak{c}/2]_q}
        }
 \label{uniQdim}
    \end{align}

\bigskip

Characters of simple Lie algebras also factorize at another point, which we call the dual Weyl vector $r$, which is a half sum over all positive coroots. Hence, we define {\bf dual quantum dimension} $\qDv_{\lambda}^{R}$ as a character at the point $q^{2r}$:
\be
     \qDv_{\lambda}^{R} : = \chi_{\lambda}^R\left(x  = q^{2r} \right)=
\prod_{\alpha \in R_+}\prod_{j=1}^{(\alpha^{\vee},\lambda)} \frac{[(\rho,\alpha^{\vee})+j]_q}{[(\rho,\alpha^{\vee})+j-1]_q}
\ee
where
\be
      r = \frac{1}{2} \sum_{\alpha\in R_+} \alpha^{\vee},\quad \alpha^{\vee} = \frac{2\alpha}{(\alpha,\alpha)}
\ee
In the case of simply laced root systems ($A_n$, $D_n$, $E_6$, $E_7$, $E_8$), the Weyl vector and the dual Weyl vector coincide $\rho = r$. In non-simply-laved cases, the dual quantum dimensions are
\be\label{B}
\qDv_{\text{Adj}}^{B_n} & =& \frac{[2n]_q[2n+1]_q}{[2]_q}, \\
\label{C}
 \qDv_{\text{Adj}}^{C_n} & =& \frac{[2n]_q[2n+1]_q}{[2]_q} \\
\qDv_{\text{Adj}}^{F_4} & =& \frac{[8]_q[12]_q[13]_q}{[4]_q[6]_q} \\
 \qDv_{\text{Adj}}^{G_2} & =& \frac{[7]_q[8]_q}{[4]_q}
\ee
In the limit of $q\to 1$, both the quantum dimensions and the dual quantum dimensions certainly reduce to the ordinary dimensions.

Note that, in variance with the ordinary dimensions, the quantum dimensions depend on the choice of normalization of the roots of algebra, and our choice of the minimal normalization is necessary for the universal formula (\ref{uniQdim}). In fact, one may say that Table 1 is written for the minimal normalization. At the same time, the dual quantum dimensions do not depend on the choice of normalization.

Despite the dual quantum dimensions also {\bf factorize}, it is unclear weather this alternative to quantum dimensions, $\qDv_{\lambda}^{R}$ can be universalized. It meets, at least, two serious problems. First of all, if there is a universal formula for the dual quantum dimensions, there should exist at once two universal formulas coinciding for the simply laced root systems and differing for the non-simply-laced ones. It is quite an intricate requirement\footnote{One can find a discussion of uniqueness of the universal formula for the quantum dimensions in \cite{AM21}.}. Second, the dual quantum dimensions for the root systems $B_n$ and $C_n$ coincide (which is a corollary of duality between the symplectic and orthogonal groups \cite{dual1,dual2,dual3,dual4}) though they are associated with completely distinct Vogel's parameters with even distinct dependence on the rank $n$. This condition is also not that simple to satisfy.

\section{Universality in refined Chern-Simons theory}

Now we make the next step and study universal quantities \cite{KS} in the refined Chern-Simons theory \cite{AgSh1,AgSh2,AM1,R,AvMkrtString,Mane,AM2,AM3}. In this case, there is a new type of dimension: the refined version of the quantum dimension and the dual quantum dimension, which we call Macdonald dimension and dual Macdonald dimension accordingly. In order to refine Chern-Simons theory with $SU(N)$ gauge group, one should go from the $sl(N)$-characters, the Schur functions to Macdonald polynomials  \cite{AgSh1, AgSh2}. Hence, it seems natural to base these refined dimensions on the Macdonald polynomials\footnote{In this letter, we use the Macdonald polynomials at values of parameters $q^2$ and $t^2$, since the universal formulas look much simpler with this choice.} $P^{(R,S)}_{\lambda}(x\,|\,t_{\alpha}^2\,|\,\,q^2,t^2)$ associated with different root systems $R$, which were introduced by Macdonald \cite{Mac}, and which provide a generalization of the famous Macdonald polynomials \cite{Mac0,Mac01}. The Macdonald polynomials associated with various root systems are symmetric w.r.t the action of the corresponding Weyl group $W_R$. They are colored with two root systems $(R,S)$ (admissible pair) and depend on parameters $q^2$ and $t^2$ as well as on parameters $t_{\alpha}^2$ that come from different Weyl orbits of the root system $R$.

Here we study the case of polynomials $P^{R}_{\lambda}(x\,|\,t_{\alpha}^2\,|\,\,q^2,t^2)$ that depend on one root system $(R,R)$, and return to mixed versions in \cite{B}.

Similarly to quantum dimensions, one naturally defines {\bf Macdonald dimension}
\be\label{MD}
\MD^R_\lambda:= P^{R}_{\lambda}\left(x = q^{2\rho_k}\,|\,t_\alpha^2\,|\,q^2,t^2\right)
\ee
with {\bf the refined Weyl vector}
\begin{equation}
    \rho_k  = \frac{1}{2} \sum_{\alpha\in R_+} k_{\alpha} \, \alpha,
    \label{refinedWeyl}
\end{equation}
where parameters $k_{\alpha}$ depend only on the length of the root $(\alpha,\alpha)$.

A problem with the Macdonald dimensions is that the Macdonald polynomial for non-simply-laced algebras does not factorize at this point, i.e. the Macdonald dimension is not factorized. For instance, the adjoint Macdonald polynomial for the root system $B_n$ evaluated at the point $q^{2\rho_k}$ is (one can find more examples, in non-adjoint representations, in \cite{B}):
\be
P^{B_n}_{\text{\text{Adj}}}\left(x = q^{2\rho_k}\,|\,t_s^2\,|\,q^2,t^2\right) = {\{t^n\}\over\{t\}}\left({\{t_s\}\over \{t\}}{\{t^{n-1}\}\over\{\xi_n\}} \{t_s\,t^{n-1}\}_{_+}\{q t^{n-2}\}_{_+}- {\{q\,t^{n-2}\}_{_+}\over\{t\}_{_+}} \frac{\{t^{n-1}\}\{t_s^2\,q^{-1}t^{-1}\}}{\{qt^{n-1}\}\{t_s\,t\xi_n\}}{_+}\right. \nn\\
\left.{_+} \frac{\{t^{n-1}\}\{t_s\}\{t\,t_s\}\{t_s \xi_n\}}{\{t^2\}\{t\xi_n\}\{\xi_n\}\{t_st\xi_n\}}
\{q^2\,t^{2(n-1)}\}_{_+}\{q^2\,t^{2(n-2)}\}_{_+} - \frac{\{q\}}{\{q\,t^{n-1}\}}
-\frac{1}{2}\{t_s t^{n-1}\}_{_+}^2 -\frac{1}{2} \frac{\{t^n\}_{_+}}{\{t\}_{_+}} \{t_s^2 t^{2(n-1)}\}_{_+}\right)
     \label{cadjnonf}
\ee
where $\xi_n:=qt_st^{2(n-2)}$. It is not that simple to universalize this kind of formulas.

\vspace{10pt}

However, I.G. Macdonald proposed \cite{Mac} a factorization formula for the Macdonald polynomials for any root system (that was later proved by I.\,Cherednik \cite{CherednikConj}) which takes place at {\bf the refined} version of the {\bf dual Weyl vector}:
\begin{equation}\label{drW}
    r_k  = \frac{1}{2} \sum_{\alpha\in R_+} k_{\alpha} \, \alpha^{\vee},
\end{equation}
where again the parameters $k_{\alpha}$ depend only on the length of root $(\alpha,\alpha)$. When $R=S$, the Macdonald factorization formula is
\begin{equation}
    P^{R}_{\lambda}\left(x = q^{2r_k}\,|\, t_{\alpha}^2\,|\, q^2, t^2 \right) = \prod_{\alpha\in R_+} \, \prod_{j=1}^{(\alpha^{\vee},\lambda)}\,  \frac{\left\{{t_{\alpha}\over q} q^{(\rho_k,\alpha^{\vee})+j}\right\}}{\left\{q^{(\rho_k,\alpha^{\vee})+j-1}\right\}}
\end{equation}
For instance, the same polynomial as in (\ref{cadjnonf}) factorizes at the refined dual Weyl vector:
\begin{equation}
     P_{Adj}^{B_n}\left(x = q^{2 r_k}\,|\,t_s^2\,|\,q^2,t^2\right) =
    \frac{[n]_t[n-1]_t}{[2]_t}
    \frac{\{t^{2n-4}\,t_s^2\}\{q\,t^{2n-4}\,t_s^2\} \{t^{2n-2}\,t_s^2\} \{q\,t^{2n-2}\,t_s^2\} }{\{t^{n-2}\,t_s\}\{q\,t^{2n-4}\,t_s\}\{t^{n-1}\,t_s\}\{q\,t^{2n-3}\,t_s\}}
\end{equation}
Similarly to the unrefined case, the Macdonald dimensions depend on the normalization of the roots of algebra, and the dual Macdonald dimensions do not.

Thus, we define {\bf dual Macdonald dimension} $\dMD_{\lambda}^{R}$ as the value of Macdonald polynomials at the point $x = q^{2r_k}$:
\begin{equation}
\dMD_{\lambda}^{R} =  P^{R}_{\lambda}\left(x = q^{2r_k}\,|\, t_{\alpha}^2\,|\, q^2, t^2 \right).
\end{equation}

Macdonald polynomials $ P^{R}_{\lambda}(x\,|\, q^2, t^2)$ associated with the simply laced root systems $R = A_n, D_n, E_6, E_7, E_8$  depend only on two parameters $q^2$ and $t^2$ and factorize at the refined Weyl vector (\ref{refinedWeyl}), since it coincides with the dual refined Weyl vector (\ref{drW}), and, hence, the Macdonald dimensions $\dMD_{\lambda}^{R}$, (\ref{MD}) in this case coincide with the dual Macdonald dimensions $\MD_{\lambda}^{R}$, (\ref{drW}).
These dimensions are equal to
\begin{equation}\label{A}
    \MD_{\text{Adj}}^{A_n} =  \frac{\{t^{n+1}\}\{q\,t^{n+1}\}}{\{t^{n}\}\{q\,t^{n}\}} \prod_{j=2}^{n} \frac{\{t^j\}}{\{t^{j-1}\}} \frac{\{t^{n-j+2}\}}{\{t^{n-j+1}\}}= \frac{\{t^n\}\{t^{n+1}\}\{q\,t^{n+1}\}}{\{t\}^2\{q\,t^{n}\}}
\end{equation}
\begin{equation}\label{D}
    \MD_{\text{Adj}}^{D_n} = \frac{ \{t^n\} \{t^{2 n-4}\} \{t^{2
   n-2}\} \{q\, t^{2 n-2}\}}{\{t\} \{t^2\}
  \{t^{n-2}\} \{q\, t^{2 n-3}\}}
\end{equation}
\begin{equation}\label{E6}
    \MD^{E_6}_{\text{Adj}} = \frac{\{t^8\}
   \{t^9\}
  \{t^{12}\} \{q\,
   t^{12}\}}{\{t\}
   \{t^3\} \{t^4\}
  \{q \,t^{11}\}}.
\end{equation}
\begin{equation}\label{E7}
    \MD_{\text{Adj}}^{E_7} =  \frac{\{t^{12}\} \{t^{14}\} \{t^{18}\}
   \{q\, t^{18}\}}{\{t\} \{t^4\}
   \{t^6\} \{q\, t^{17}\}}.
\end{equation}
\begin{equation}\label{E8}
     \MD_{\text{Adj}}^{E_8} = \frac{\{t^{20}\}\{t^{24}\}\{t^{30}\}
   \{q\, t^{30}\}}{\{t\}\{t^6\}
  \{t^{10}\} \{q\, t^{29}\}}.
\end{equation}

In the case of non-simply-laced root systems $R = B_n,C_n,F_4,G_2$, the Macdonald polynomials $P^{R}_{\lambda}(x\,|\,t_{\alpha}^2\,|\,q^2,t^2)$  depend on an additional parameter $t_{\alpha}^2$ associated with the root of a distinct length.  Since, in the non-simply-laced case, there are a few deformation parameters $t^2$, $t_{\alpha}^2$, and, in the simply laced case, there is just one $t^2$, one does not have to expect that there is a universality in the generic case. However, in the simply laced case, the universality still persists.

There is another argument against universality of all algebras after the refinement. It is connected to the fact that the quantum dimensions celebrate universality, while the dual quantum dimensions seem not to universalize as we discussed in the previous section. Hence, it is expected that their refined versions do not universalize as well.

At the same time, the adjoint Macdonald polynomials associated with the simply laced root systems factorize according to formulas (\ref{A}), (\ref{D}), (\ref{E6}), (\ref{E7}), (\ref{E8}) and can be unified with a universal formula which one can call the simply laced universal Macdonald dimension (which coincides with the dual one):

\begin{equation}\label{main1}
\boxed{
    \MD_{\text{Adj}}(\mathfrak{a},\mathfrak{b},\mathfrak{c}) =    - {\{t^{\mathfrak{a}+\mathfrak{b}/2+\mathfrak{c}}\}\{t^{\mathfrak{a}+\mathfrak{b}+\mathfrak{c}/2}\}
    \{t^{\mathfrak{a}+\mathfrak{b}+\mathfrak{c}}\}\{q\,t^{\mathfrak{a}+\mathfrak{b}+\mathfrak{c}}\}\over
    \{t^{\mathfrak{a}/2}\}\{t^{\mathfrak{b}/2}\}
    \{t^{\mathfrak{c}/2}\}\{q\,t^{\mathfrak{a}+\mathfrak{b}+\mathfrak{c}-1}\}}
    }
\end{equation}
There is also a more symmetric form:
\begin{equation}\label{main2}
    \boxed{\MD_{\text{Adj}}(\mathfrak{a},\mathfrak{b},\mathfrak{c}) =  - \frac{\{t^{\mathfrak{a}/2+\mathfrak{b}+\mathfrak{c}}\}\{t^{\mathfrak{a}+\mathfrak{b}/2+\mathfrak{c}}\}\{t^{\mathfrak{a}+\mathfrak{b}+\mathfrak{c}/2}\}
    \{t^{\mathfrak{a}+\mathfrak{b}+\mathfrak{c}}\}\{q\,t^{\mathfrak{a}+\mathfrak{b}+\mathfrak{c}}\}}{\{t^{\mathfrak{a}/2}\}
    \{t^{\mathfrak{b}/2}\}\{t^{\mathfrak{c}/2}\} \{t^{\mathfrak{a}+\mathfrak{b}+\mathfrak{c}+1}\} \{q\,t^{\mathfrak{a}+\mathfrak{b}+\mathfrak{c}-1}\}}}
\end{equation}

This universality formula for the simply laced algebras is the main result of the paper.
It is in accordance with earlier claims by authors of \cite{KS,AM1,Mane}, who studied the refinement of the Chern-Simons partition function on $S^3$ and also realized that universality holds for the simply laced algebras.

\section{Comments on non-simply-laced root systems}

Note that the dual Macdonald dimension in the $B_n$ case is equal to that in the $D_n$ case at $t_s=1$, $\MD_{\text{Adj}}^{D_n}$ \cite{ts=1}. Moreover, the ordinary Macdonald dimension (\ref{cadjnonf}) drastically simplifies and factorizes at $t_s=1$ and coincides with $\MD_{\text{Adj}}^{D_n}$. However, the two root systems $B_n$ and $D_n$ are described by distinct Vogel's parameters, hence, the $B_n$ series still does not match the universality formula at $t_s=1$.

The reason for coinciding dimensions is that there are duality relations \cite{ts=1,B} (the polynomials here are evaluated at the normalization\footnote{We remind that the dual Macdonald dimensions do not depend on normalization.} where for all roots but may be one $(\alpha,\alpha)$=2)
\be
P^{B_n}_{[1,1]}\left(x\,|\,1\,|\,q^2,t^2\right)=P^{D_n}_{[1,1]}\left(x\,|\,q^2,t^2\right)=P^{C_n}_{[1,1]}\left(x\,|\,1\,|\,q^2,t^2\right)
\ee
In fact, these relations can be extended to other representations (note that $[1,1]$ is the adjoint representation for $B_n$ and $D_n$, and $[2]$, the adjoint representation for $C_n$). However, for constructing universality formulas, this kind of relations is of no use since the Macdonald polynomials at $t_{s,l}=1$ do not match at $t=q$ the corresponding Lie group characters and, hence, the quantum dimensions.

At last, note the Macdonald dimension in the $B_n$ case (\ref{cadjnonf}) at $t_s=q$ also factorizes and takes an intriguing form
\be\label{27}
P^{B_n}_{\text{\text{Adj}}}\left(x = q^{2\rho_k}\,|\,t_s^2\,|\,q^2,t^2\right)\Big|_{t_s=q} ={\{t^{n-1}\}\{t^n\}\{q^{3\over 2}t^{n-1}\}\{q^{3\over 2}t^{n-2}\}\{q^2t^{2(n-1)}\}\{qt^{2(n-2)}\}\{qt^{2(n-1)}\}\over
\{qt^{n-1}\}\{q^{1\over 2}t^{n-2}\}\{q^{1\over 2}t^{n-1}\}\{t^2\}\{qt^{n-2}\}\{q^2t^{2n-3}\}\{t\}}=\nn\\
={\{t^{n-1}\}\{t^n\}\{q^{3\over 2}t^{n-1}\}\{q^{3\over 2}t^{n-2}\}\over
\{t^2\}\{qt^{n-2}\}\{q^2t^{2n-3}\}\{t\}}\{qt^{n-1}\}_{_+}\{q^{1\over 2}t^{n-2}\}_{_+}\{q^{1\over 2}t^{n-1}\}_{_+}
\ee
This formula does not look matching any universal formulas for dimensions of the Macdonald polynomials for other root systems. However, it implies that, in the refined case, there might be yet another parameter that inclines the straight lines corresponding to the non-simply-laced algebras from the Vogel's parameter plane to $3d$ Euclidean space.

Still, in the case of root system $C_n$, the factorization formula is more involved:
\be\label{28}
P^{C_n}_{\text{\text{Adj}}}\left(x = q^{2\rho_k}\,|\,t_l^2\,|\,q^2,t^2\right)\Big|_{t_l={1\over\sqrt{qt^{n-1}}}} =
{\{t^{n\over 2}\}\{t^{n-1\over 2}\}\{q^{1\over 2}t^{n\over 2}\}\{q^2\}\{tq^{-1}\}\over\{q^{1\over 2}t^{n-1\over 2}\}\{t^{1\over 2}q^{-{1\over 2}}\}
\{q^{1\over 2}\}\{t^{1\over 2}\}\{qt\}}
\ee
in particular, the factorization point is now $n$-dependent. The main drawback of this is that such $t_l$ does not reduce to $q$ at $t=q$ that provides a factorized expression for the quantum dimension (a particular case of the universal quantum dimension (\ref{uniQdim})) in the unrefined case.

The technical reason for the $C_n$ case to look more complicated is that, in the minimal normalization (which is associated with the Vogel's universality), the answers are far more complicated\footnote{For instance, a counterpart of formula (\ref{28}) in the standard normalization, when, for the long root of the root system $C_n$, $(\alpha_l,\alpha_l)=4$ is
\be
\overline{P}^{C_n}_{\text{\text{Adj}}}\left(x = q^{2\rho_k}\,|\,t_l^2\,|\,q^2,t^2\right)\Big|_{t_l=q} =
{\{t^n\}\{qt^n\}\{q^3t^{2n-2}\}\{q^6t^{2n-2}\}\over\{q^2t^{n-1}\}\{q^3t^{n-1}\}\{t\}\{qt\}}
\ee
and one can see, it is again factorized at the point $t_l=q$ as in the $B_n$ case.}, and the minimal normalization makes a difference only in the $C_n$ and $G_2$ cases.

Another distinguished point in the $C_n$ case is $t_l=t$. At this point, the answer is not factorized but can be presented in a relatively simple form:
\be
P^{C_n}_{\text{\text{Adj}}}\left(x = q^{2\rho_k}\,|\,t_l^2\,|\,q^2,t^2\right)\Big|_{t_l=q} &=&{\{t^{n\over 2}\}\over\{qt\}\{t\}}\Biggl(
\{t^{n+1\over 2}\}_{_+}\{qt^{n+1}\}+\{t^{1\over 2}\}_{_{+}}\{qt^{3n+1\over 2}\}+\nn\\
&+&\{t^{n-1\over 2}\}\left(\{qt^{n\over 2}\}+
\{t^{n+1\over 2}\}_{_+}\{q^{-1}t^{{n\over 2}+1}\}_{_+}\right)\Biggr)
\ee
An advantage of this point is that it conveniently reduces to the proper quantum dimension at $t=q$.

Full details of the status and justification of formulas of this and the previous sections and a detailed and extended discussion can be found in \cite{B}.

\section{Conclusion}

In this letter, we extended the Vogel's universality, which is a feature of Chern-Simons theory, to the refined Chern-Simons theory with gauge groups $A_n$, $D_n$ and $E_6$, $E_7$, $E_8$, i.e. those associated with the simple laced root systems. The quantity that has a universal description is the Macdonald dimension, which is the value of the Macdonald polynomial
associated with the corresponding root system at the point $q^{2\rho}$. This quantity can be presented in a universal form (\ref{main1})-(\ref{main2}).

Note it was earlier observed \cite{KS,AM1,Mane} that the partition function of the refined Chern-Simons theory can be also presented in a universal form in the case of simple laced root systems. This means that one can naturally expect an extension of all previously known universal quantities to the refined case in the case of simple laced root systems, all of them being basically related to knot theory.

In particular, one expects to construct universal hyperpolynomials of knots and links, which generalize the universal HOMFLY-PT polynomials \cite{MMM,MM} and are hyperpolynomials of knots/links in the adjoint representations. In the case of link, all the link components have to be in the adjoint representation.

A particular case of these knot/link polynomials is the hyperpolynomial of unknot, which is given exactly by the Macdonald dimension. Hence, the universal formula for unknot has been already constructed in this letter. The next goal is to construct a universal formula for less trivial knots/links, the simplest of them are the trefoil knot and the Hopf link. In fact, the hyperpolynomial of the Hopf link is known \cite{MMPHopf} to be described by the Cherednik-Macdonald-Mehta formulas \cite{Che,EK,ChaE}, and constructing its universal form would open a new direction for the Vogel's universality. We will return to these issues elsewhere.

Note that formulas like (\ref{27}), (\ref{28}) imply that, in the refined case, it may be reasonable to add one more parameter to the Vogel's parameters, which describes deviation from the Vogel's plane of the non-simply-laced algebras. Another issue is including the root systems $(R,S)$ at $R\ne S$ into the Vogel's framework. We will return to these two issues elsewhere.

More possible implications of the Vogel's universality can be found in \cite{HMS}.

\section*{Acknowledgements}

This work is supported by the RSF grant 23-71-10058.

\end{document}